\documentclass[journal]{IEEEtran}
\usepackage{amssymb,amsmath,amsfonts,amsthm}
\usepackage{mathrsfs}
\usepackage{cite}
\usepackage{array}
\usepackage{tabularx}
\usepackage[table]{xcolor}
\usepackage{color}
\usepackage{mathtools}
\usepackage{subfigure}
\usepackage{mathtools}
\usepackage{graphicx}
\usepackage{bbm}
\usepackage[utf8]{inputenc}

\usepackage{multirow}
\usepackage{float}

\newtheorem{definition}{Definition}

\title{Throughput and Delay Analysis of Wireless Caching Helper Systems with Random Availability}

\author{\IEEEauthorblockN{Nikolaos Pappas~\IEEEmembership{Member,~IEEE}, Zheng Chen~\IEEEmembership{Member,~IEEE}, Ioannis Dimitriou
\thanks{N. Pappas is with the Department of Science and Technology, Link\"{o}ping University, Norrk\"{o}ping SE-60174, Sweden. (e-mail: nikolaos.pappas@liu.se).}
\thanks{Z. Chen is with Department of Electrical Engineering,
Link\"{o}ping University, Link\"{o}ping, Sweden. (e-mail: zheng.chen@liu.se).}
\thanks{I. Dimitriou is with the Department of Mathematics, University of Patras, Patra, Peloponnese, Greece.
(e-mail: idimit@math.upatras.gr).}
}}

\begin{document}

\maketitle

\begin{abstract}
In this paper, we investigate the effect of bursty traffic and random availability of caching helpers in a wireless caching system. More explicitly, we consider a general system consisting of a caching helper with its dedicated user in proximity and another non-dedicated user requesting for content. Both the non-dedicated user and the helper have limited storage capabilities. When the user is not able to locate the requested content in its own cache, then its request shall be served either by the caching helper or by a large data center.  
Assuming bursty request arrivals at the caching helper from its dedicated destination, its availability to serve other users is affected by the request rate, which will further affect the system throughput and the delay experienced by the non-dedicated user. 
We characterize the maximum weighted throughput and the average delay per packet of the considered system, taking into account the request arrival rate of the caching helper, the request probability of the user and the availability of the data center. Our results provide fundamental insights in the throughput and delay behavior of such systems, which are essential for further investigation in larger topologies. 
\end{abstract}


\section{Introduction}
Driven by the development of information-centric applications, wireless caching has emerged as a promising concept to cope with the exponential growth of data traffic, of which video content is the dominant source. Caching at the network edge also exploits the high degree of asynchronous content reuse among users who share similar content preferences in local areas. By introducing caching capabilities at the network edge, including small cells, femto access points and user devices, and caching popular content closer to end users before being requested, cellular traffic load and latency can be significantly reduced \cite{femtocaching, paschos_commag, Niklas2017, chen_d2d}.    

In the literature of wireless caching, various types of content placement strategies have been investigated and discussed, such as ``cache the most popular content everywhere", random/probabilistic caching \cite{optimalcaching, tradeoff_cooperation}, and cooperative caching\cite{chen_cooperation, MaAccess2017}. Based on specific caching schemes under consideration, different network performance metrics have been studied and optimized, such as the cache hit ratio, the cache-aide throughout \cite{Zheng-comml2016}, the energy efficiency \cite{ee_caching}, etc.
A cooperation-based scheme has been proposed in \cite{MaAccess2017} which aims at minimizing the average energy consumption incurred by a user equipment in order to obtain its desired content. 
In \cite{WangAccess2017}, the transmission cost in edge caching has been minimized by taking into account traffic offloading via device-to-device communications. The users’ quality of experience (QoS) is considered in \cite{TanzilAccess2017} with an algorithm being proposed to improve the user QoS and reduce the network traffic using machine learning approach. The algorithm considers users’ behavior and properties of the cellular network.
In \cite{LiAccess2017}, the concept of ‘‘Caching-as-a-Service" has been proposed based on cloud radio access networks, and virtualized evolved packet core, which provides reliable and flexible cache service with high elasticity and adaptivity.
In \cite{XWangAccess2017}, social-aware edge caching techniques in fog radio access networks have been investigated by modeling the impact of edge caching schemes on the performance of content diffusion and sharing.

\subsection{Related work}
In the early studies of wireless caching systems, the cache hit probability and the density of successful receptions have been commonly used to evaluate the performance of certain wireless caching systems/schemes. Recently, there is growing interest in the delay analysis or the combination of throughput and delay analysis in cache-enabled wireless networks, which extend the research in this area towards another perspective \cite{rate_delay_coded, delay_geographical, caching_user_delay}. Nevertheless, most of the delay analysis in the literature focuses on either the backhaul delay or the transmission delay under the saturated traffic/requests assumption. Based on queueing theory, the consideration of bursty traffic model provides new insights that cannot be easily seen with the saturated traffic assumption \cite{queueing_it}. In \cite{Rezaei-TCOM2016}, the stable throughput and delay performance of single bottleneck caching networks have been studied by considering stochastic arrivals of the requests at different nodes. 

Another common assumption in the studies of wireless caching systems is that a caching helper will be able to serve a user's request whenever it has the requested file cached. In reality, a caching helper might not be available to assist the requests of nearby users when it has other destinations to serve. For instance, in a FemtoCaching network where the Femto Base Station (FBS) has some dedicated users that are requesting content in a random manner, another random user falling inside its coverage area will probably not be served immediately even when the FBS has its requested file cached inside.
The bursty traffic at a caching helper can affect its availability and the response time to the non-dedicated user waiting to be served. In this case, the throughput and the delay depend not only on the downloading delay which is often related to the channel condition, but also on the response time from the caching helper whose availability is affected by the request arrival rate of its dedicated users.

\subsection{Contributions}
In this paper, we investigate the effect of bursty traffic on the throughput and the delay performance of a wireless caching helper network. We consider the case where a source helper with limited cache storage has bursty traffic to transmit to its intended destination, and the source stores the traffic to a queue. In addition, there is another user requesting for content, the user has also limited storage capability. If the content is not located in its cache, then the user will seek for assistance from either the helper or a distant data center. However, since the helper has also limited storage capability a cache miss can occur even if it is available to assist the user. The data center contains all the files that can be requested by the user but is not always available due to congestion.

Based on a random access scheme between the user, the caching helper and its dedicated destination, we characterize the performance of this network in terms of throughput and delay. More specifically, we characterize the network-wide throughput in both cases where the queue at the source helper is stable and unstable. In addition, we optimize the maximum weighted network throughput having stability constraints for the source/helper. Furthermore, we derive the average delay seen by the user when is requesting for content that cannot be located in its storage. Finally, we provide numerical evaluation of the presented results.

Our analysis builds on a simple network model with four nodes, but the general conditions of random access probabilities can capture many specific types of caching network structures and different access criteria related to the channel condition, the request load and the backhaul availability etc. The analysis here can be used for further investigation in larger topologies. To the best of our knowledge, similar results to this work have not been reported in the literature.

\subsection{Organization of the paper}
In Section \ref{sec:model}, we present the considered system model including the network model, the caching policies, the transmission model, and the physical layer considerations. In Section \ref{sec:thr}, the analytical results regarding throughput are derived including the cases of stable and unstable queue at the helper. The average delay performance is derived in Section \ref{sec:delay}. In Section \ref{sec:results}, we evaluate numerically the theoretical results. Finally, we conclude our paper in Section \ref{sec:conclusions}.

\section{System Model} \label{sec:model}
\subsection{Network Model}
\begin{figure}[ht!]
	\centering
	\includegraphics[scale=0.4]{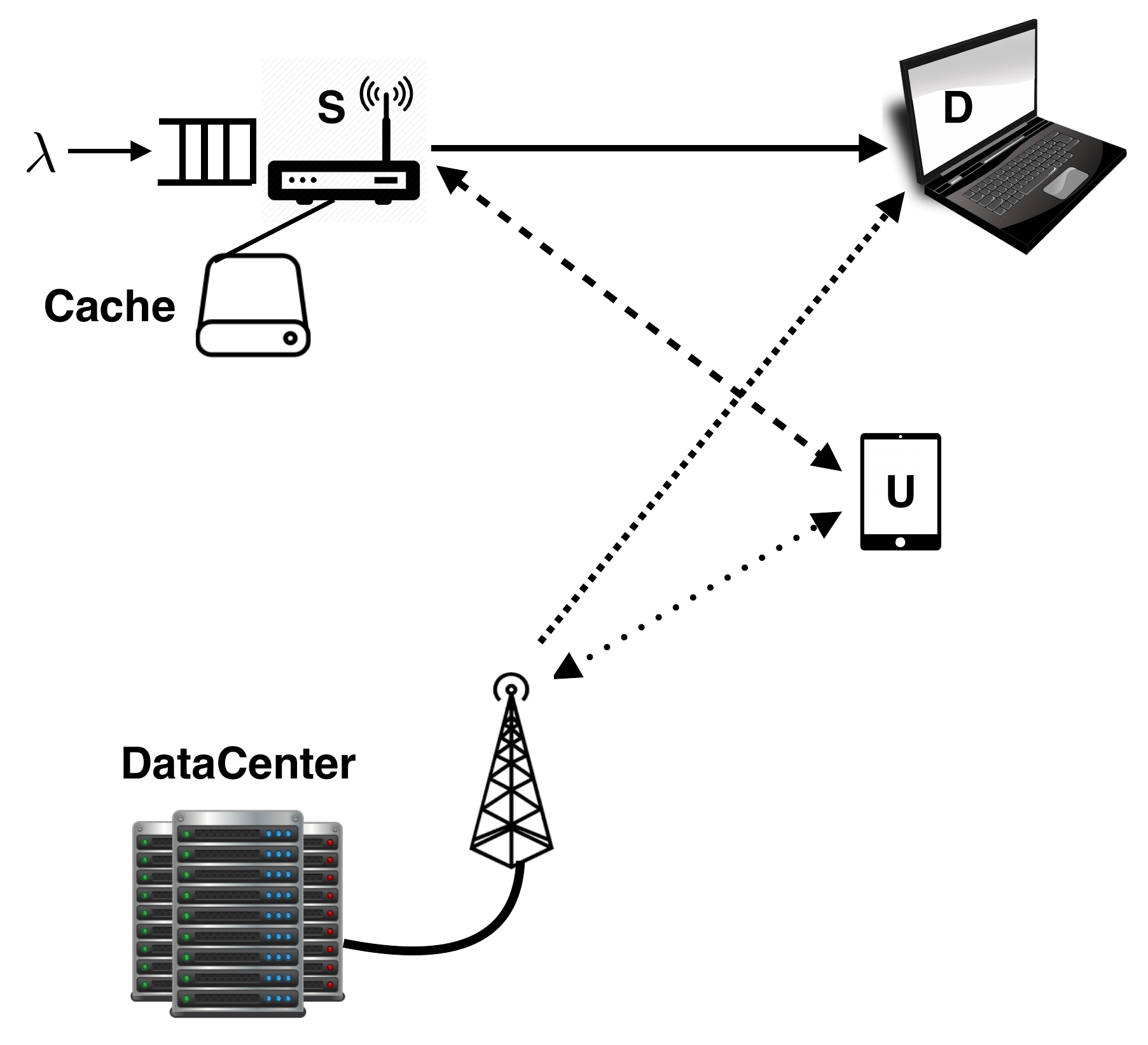}
	\caption{An example of our system model. The caching helper $S$ can be a femto access point (FAP) with storage capabilities. It has a dedicated connected user $D$, who is randomly generating requests for multimedia content. At the same time, there is a mobile device $U$ that is in proximity with $S$, which requests for external content with some probability in each time slot. The device $U$ also has access to the data center $DC$, but the connection is problematic due to the congestion, so it is preferred to be served by $S$ when possible.}
	\label{fig:model}
\end{figure}

We consider the following network model: a caching helper $S$ has bursty packets that need to be transmitted to its dedicated user $D$. The helper $S$ is equipped with an infinite queue size and the packet arrival at $S$ is modeled by a Bernoulli process with average arrival rate $\lambda$. In this study, we assume slotted time and the transmission of a packet requires one timeslot. We assume another non-dedicated random user $U$ that falls inside the service range of the caching helper, as depicted in Fig. \ref{fig:model}.  In each timeslot, the user $U$ requests for a file, with probability $q_U$ cannot locate this file in its storage, thus $U$ will request the file externally.
The helper $S$ under certain conditions will assist the requests of $U$. Explicitly, in each timeslot, if the queue at $S$ is not empty, $S$ will transmit to $D$ with probability $q_S$, and with probability $1-q_S$ it will be available to assist $U$. When $S$ is not transmitting to $D$, including the cases with non-empty and empty queues, $S$ will assist the request of $U$ with probability $q_C$.

Note that the probabilities considered in this model can cover many different aspects of uncertainty/variation in a wireless caching system. As mentioned, the probability $q_U$ describes the activity of the user in terms of how often the user is requesting for \textit{external} content. This can include the case where the user has some limited storage capabilities. Then, $q_U$ is related to the probability that the requested content is not located in its storage and needs to be delivered from external sources. Thus, $q_U=1$ denotes the case where the user does not have storage capabilities and has to seek for its content either from the helper or the data center, another case that can be represented by $q_U=1$ is when the cache of $U$ contains obsolete/outdated content. If $q_U=0$, then all the requested content by the user is located in its storage unit thus, does not need assistance by neither the helper nor the data center. More details will be given in Section \ref{sec:caching_policy}. The probability $q_S$ can be related to the random access probability that potentially needs to be optimized in the case with many concurrent caching helpers sharing the same frequency/time resources. The probability $q_C$ might come from the cache updating process inside the caching helper. The distinction between dedicated and non-dedicated users can capture the possibility of having users with different priorities.

Recall that when the non-dedicated user $U$ is requesting for a file, it has probability $q_C$ to be served by $S$ when available. With probability $1-q_C$ the request will be directed to a large data server denoted by $DC$, which is assumed to have every available content stored. As a result, in a timeslot, the probability that the user $U$ will request for a file and the request will be served by $S$ is $q_U (1-q_S) q_C$, when the queue at $S$ is not empty. Due to the limited storage capacity of the helper $S$, the requested file of user $U$ has probability $p_h$ to be cached within the helper, thus, there is a probability $p_m=1-p_h$ that $U$ cannot find the requested content in $S$ and has to request it from $DC$. Here, $p_h$ depends on the caching strategy and the user's request pattern.

The operation of the caching helper $S$ is summarized as a flowchart in Fig. \ref{fig:flowchartS}. The operation of the device $U$ regarding the possible cases of finding the requested content is given in Fig. \ref{fig:flowchartU}. The operation of the considered system is rather complicated even in such a simple topology.

\subsection{Request Distribution and Caching Policy} \label{sec:caching_policy}
We consider a finite content library $\mathcal{F}=\{f_1, \cdots, f_N\}$ for the user requests, where $f_i$ is the $i$-th most popular file and $N$ is the library size. All files are assumed to have equal size, which is normalized to one. We use the standard Zipf law for the popularity distribution, meaning that the request probability of the $i$-th most popular file is 
\begin{equation}
p_{i}=\frac{\Omega}{i^\delta},
\end{equation}
where $\Omega=\left(\sum\limits_{j=1}^{N}j^{-\delta}\right)^{-1}$ is the normalization factor and $\delta$ is the shape parameter of Zipf law, which defines the correlation level of user requests. High values of $\delta$ means that most of the requests are generated from a few most popular files. For a user making a random request, $p_i$ can be seen as the probability that the requested file is $f_i$.

We assume that the user device $U$ and the caching helper have cache storage capacity equal to $M_U$ (files) and $M_S$ (files), respectively with $M_S \geq M_U$.

In the literature of content caching in wireless networks, several proactive caching policies have been used.
The policy \textit{cache the Most Popular Content} (MPC) policy, where each node independently stores the files with the highest popularity, has been extensively used in the literature. The MPC strategy gives optimal performance with non-overlapping SBS coverage or with isolated caches. 
With this strategy, the user stores the $M_U$ most popular files in its own cache and the caching helper stores the $M_S$ most popular files. The probability for user $U$ to request for a file that is not cached within its own cache is:
\begin{equation} \label{eq:qU}
q_{U}=1-\sum\limits_{i=1}^{M_U} p_{i}.
\end{equation}
The cache hit probability at the caching helper is
\begin{equation} \label{eq:phMPC}
p_{h}=\sum\limits_{i=M_U+1}^{M_S} p_{i}.
\end{equation}

Since the nodes store independently the most popular files, in case of a miss at $U$, which happens with probability $q_U$, then requested content will not be among the first $M_U$ files at $S$.

We will aslo consider a variation of MPC, the collaborative MPC (CMPC), where the user $U$ stores the first $M_U$ most popular files, then the helper $S$ knowing this will store the next $M_S$ most popular files. In this case is similar to merging the two storage elements into one. The probability for user $U$ to request for a file that is not cached within its own cache is given by \eqref{eq:qU}. The cache hit probability at the caching helper is

\begin{equation} \label{eq:phCMPC}
p_{h}=\sum\limits_{i=M_U+1}^{M_S+M_U} p_{i}.
\end{equation}

However, CMPC requires some exchange of information between the devices such as the storage size and the collaborative decisions of content placement in the device and the helper. 

\begin{figure}[t]
\centering
 \subfigure[$\delta=0.5$.]{
 \includegraphics[scale=0.47]{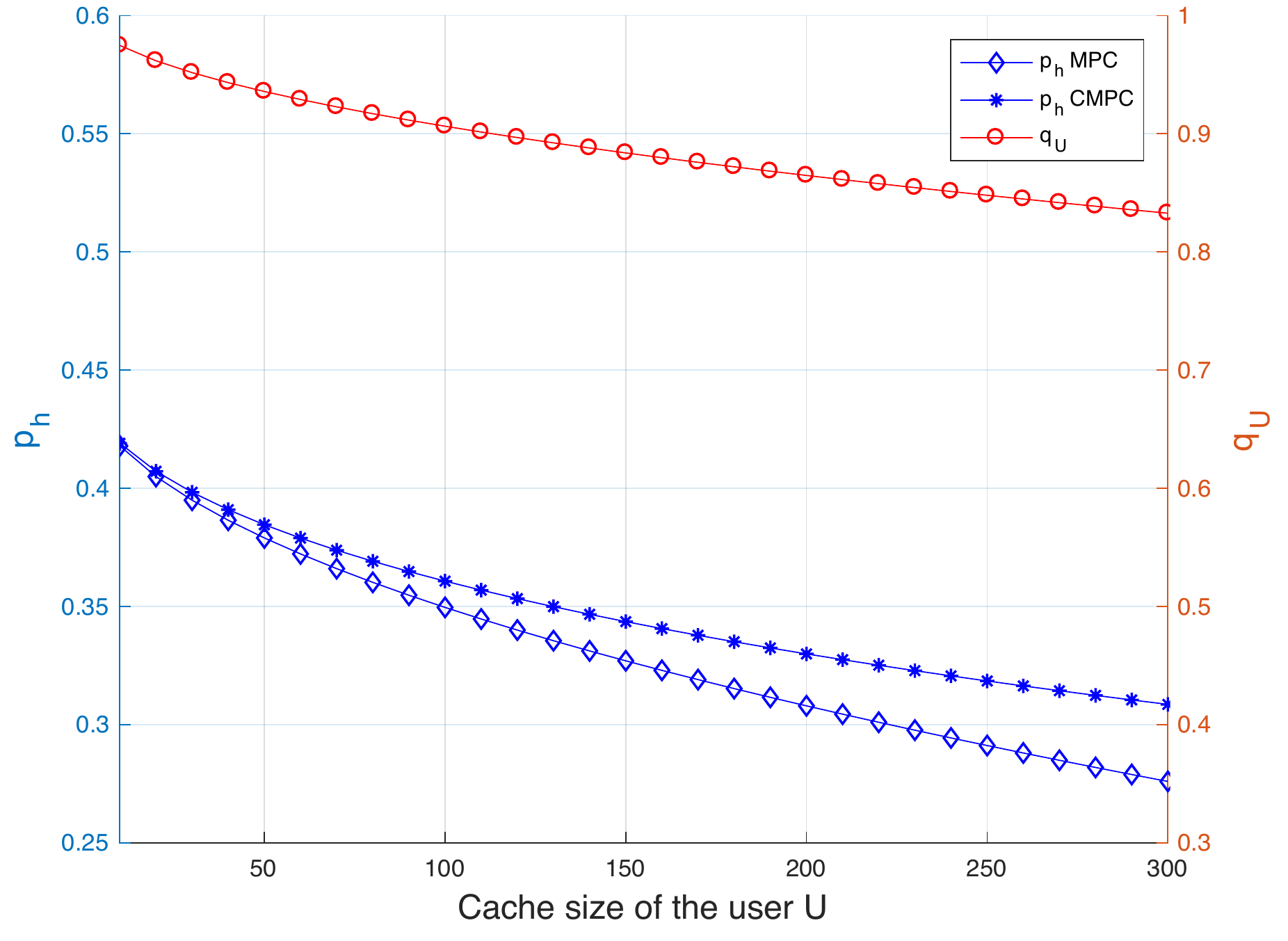}
 \label{fig:ph1}
 }

  \subfigure[$\delta=0.9$.]{
  \includegraphics[scale=0.47]{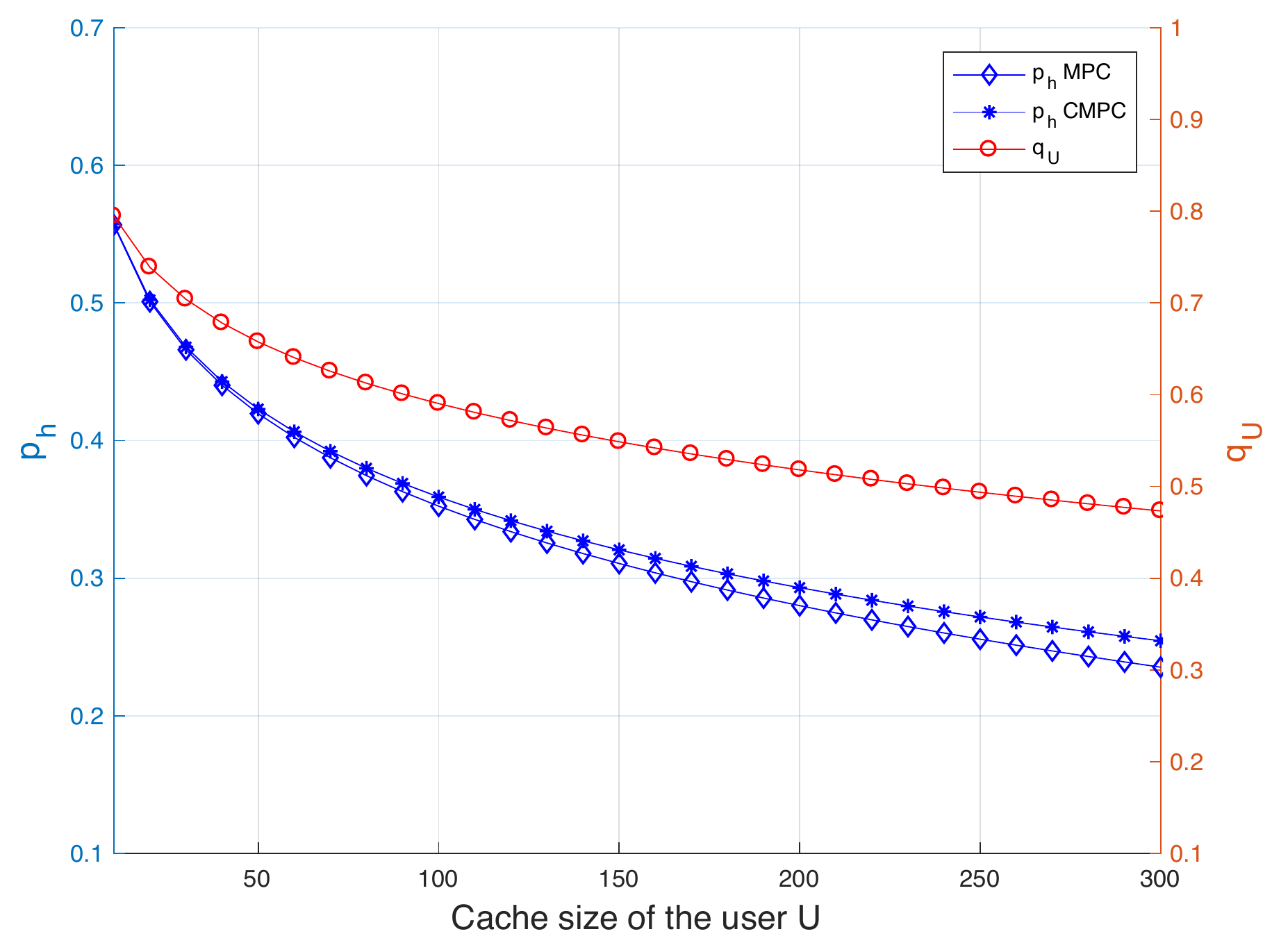}
  \label{fig:ph2}
  }
  \caption{We depict the cases where the MPC and CMPC schemes are applied for several values of $M_U$ in a scenario where the content library contains $10000$ files, the capacity of the cache at source is $M_S=2000$, and the shape parameter of the Zipf law. On the left y-axis we have the values of $p_h$ on the right y-axis we have the values of $q_U$}\label{fig:ph}
\end{figure}

In Fig. \ref{fig:ph}, we evaluate the cache hit probability at the caching helper and the probability that the user $U$ will request for a file that cannot be found in its cache.

In terms of $p_h$ the difference is small, thus, in the remainder of this paper we will consider the MPC scheme.

\textit{The results obtained in this work are general and they hold for all the possible request distributions and caching policies as long as one can replace the $q_U$ and $p_h$ with the appropriate expressions.}

\subsection{Transmission Model}
\begin{figure*}[ht!]
	\centering
	\includegraphics[width=\textwidth]{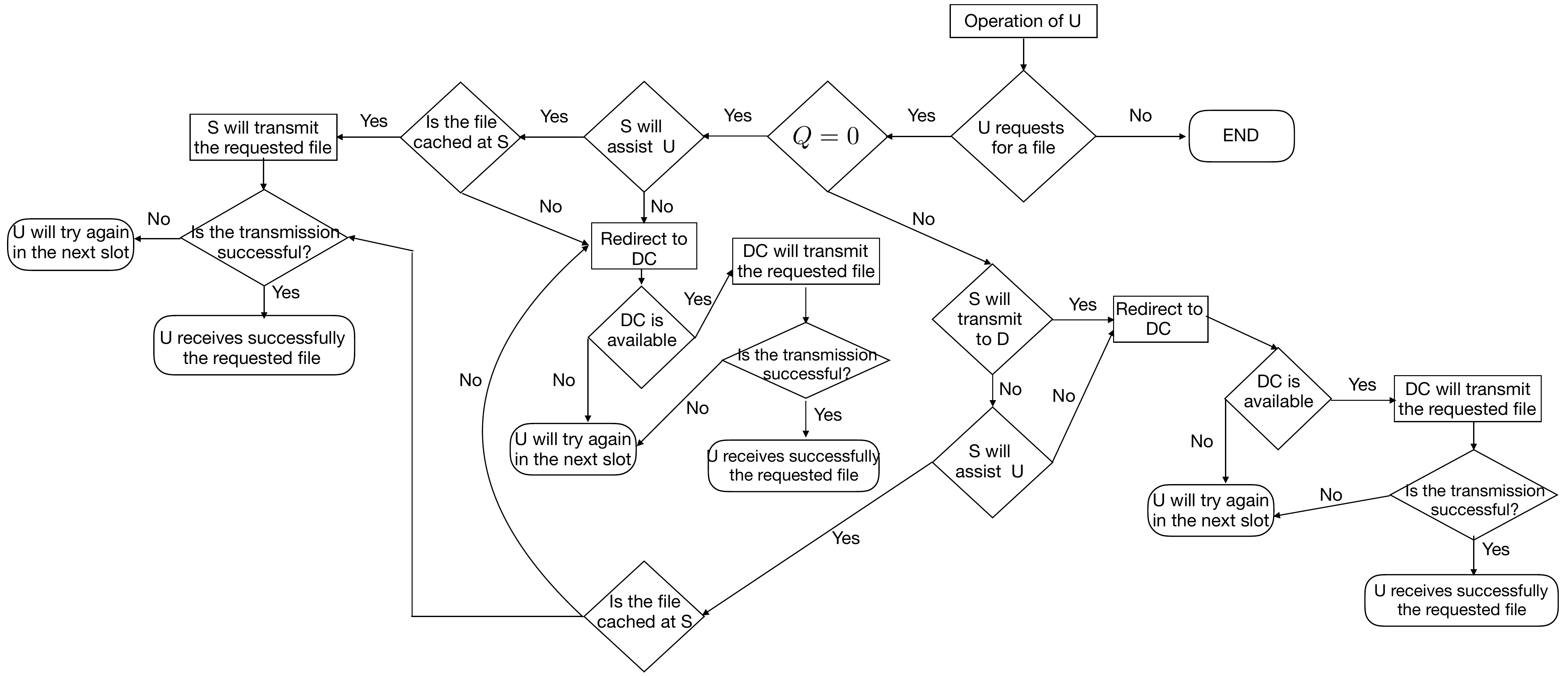}
	\caption{The operation of user $U$ in the described protocol when $U$ is unable to locate the requested content in its cache.}
	\label{fig:flowchartU}
\end{figure*}

\begin{figure*}[ht!]
	\centering
	\includegraphics[width=\textwidth]{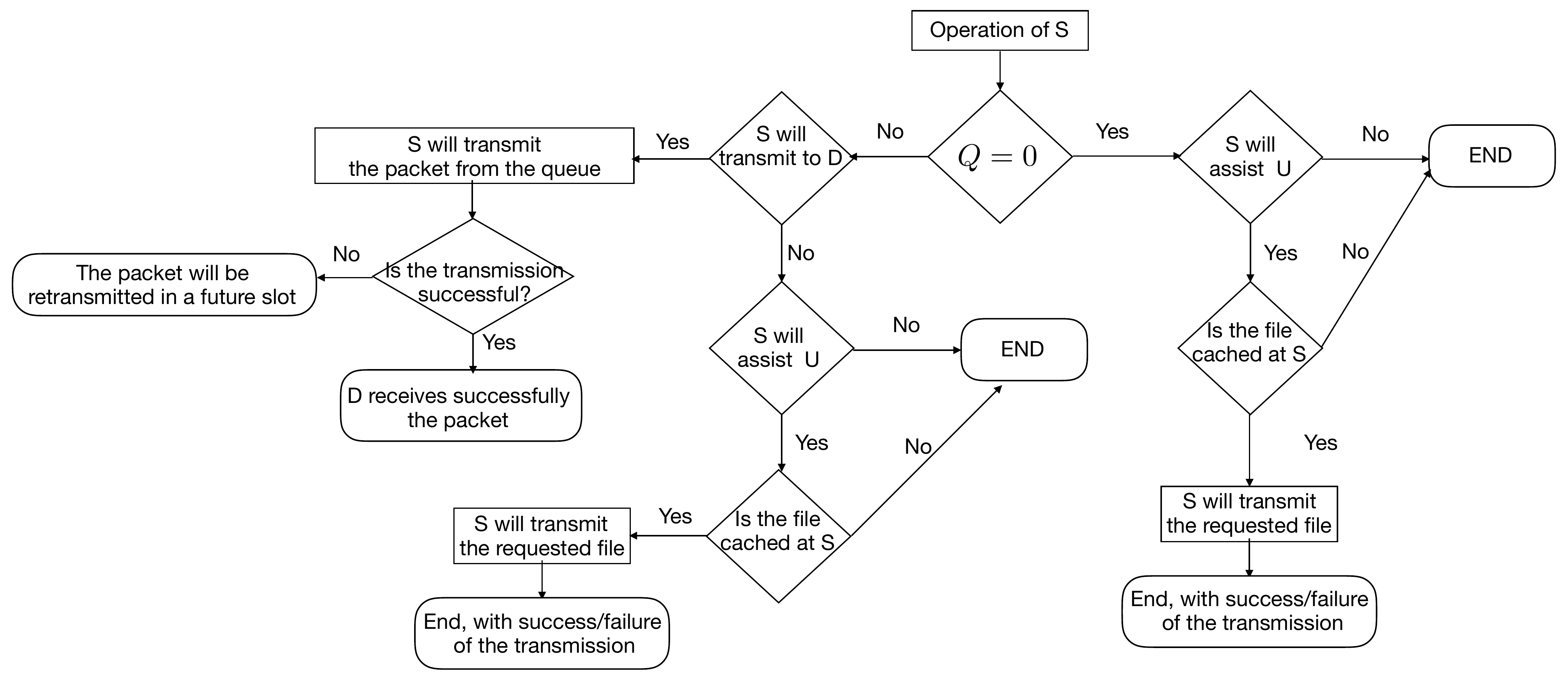}
	\caption{The operation of source $S$ in the described protocol.}
	\label{fig:flowchartS}
\end{figure*}

When the caching helper $S$ is transmitting to its destination $D$, the user $U$ will seek its requested information directly from the $DC$. If $DC$ is available to serve $U$, then there will be two parallel transmissions, one from $DC$ to $U$ and the other from $S$ to $D$, which are interfering. Considering that $DC$ might be congested with other users' requests, we model the availability of $DC$ with a factor $\alpha$, which models the probability that $DC$ is available in a timeslot to serve $U$. If $\alpha=0$, then either $DC$ does not exist in the network or it is heavily congested and thus not available for $U$.
If $DC$ is always available to the user, then $\alpha=1$. The aforementioned probabilities are summarized in Table \ref{table:notation}. 
Figs. \ref{fig:flowchartU} and \ref{fig:flowchartS} provide flowcharts describing the operation of $U$ and $S$ respectively.

\begin{table}[!t]
	\renewcommand{\arraystretch}{1.6}
	\caption{Notation Table}
	\centering
	\begin{tabular}{ | c | l | }
		\hline
		\bfseries Probability & \bfseries Description\\
		\hline
		$q_U$ & $U$ will request for content outside its storage\\
		$q_S$ & $S$ will transmit to $D$ if its queue is not empty\\
		$q_C$ & $U$ will be assisted by the cache of $S$ if $S$ is available\\
		$p_h/p_m$ & Cache hit/miss probability from the cache of $S$\\
		$a$ & $DC$ is available to serve $U$\\ 
		\hline
	\end{tabular} \label{table:notation}	
\end{table}

\subsection{Physical Layer Model}
We denote $p_{SD/S}$ the success probability of the link $S \rightarrow D$ when only $S$ is active, and $p_{SD/S,DC}$ is the success probability of the link $S \rightarrow D$ when both $S$ and $DC$ are transmitting. We consider the success probability of each link $i\rightarrow j$ based on its received signal-to-interference-plus-noise ratio (SINR)
\begin{equation*}
{\rm SINR}_{ij}=\frac{P_{i}|h_{i,j}|^2 r_{i,j}^{-\gamma}}{\eta_j+\sum_{k\in T\backslash\left\{i\right\}} P_{k}|h_{k,j}|^2 r_{k,j}^{-\gamma}},
\end{equation*}
where $T$ denotes the set of active transmitters; $P_{i}$ denotes the transmission power of node $i$; $h_{i,j}$ denotes the small-scale channel fading from the transmitter $i$ to the receiver $j$, which follows $\mathcal{CN}(0,1)$ (Rayleigh fading); $r_{i,j}$ denotes the distance from the transmitter $i$ to the receiver $j$; $\eta_j$ denotes the background thermal noise power received at $j$. 

Denote $p_{ij/i}$ the success probability of link $i\rightarrow j$ when only transmitter $i$ is active, we have
\begin{equation}\label{eq:succprobSNR}
p_{ij/i} =\mathbb{P} \left\lbrace \mathrm{SNR}_{ij} \geq \theta_j \right\rbrace = \exp \left(- \frac{\theta_j \eta_j r^{\gamma}_{ij}}{P_{i}}\right).
\end{equation}
Similarly, denote $p_{ij/i,k}$ the success probability of link $i\rightarrow j$ when both transmitters $i$ and $k$ are active, thus we have
\begin{align}\label{eq:succprobSINR}
p_{ij/i,k}=\mathbb{P} \left\lbrace \mathrm{SINR}_{ij} \geq \theta_j \right\rbrace = \frac {\exp \left(- \frac{\theta_j \eta_j  r^{\gamma}_{ij}}{P_{i}} \right)}{ \left[1+\theta_j \frac{P_{k}}{P_{i}} \left( \frac{r_{ij}}{r_{kj}} \right)^\gamma \right]}.
\end{align}
The above expression for the success probabilities are obtained with Rayleigh fading. The results obtained in this work hold for other wireless channels as well. One needs to connect the expressions for the success probabilities and then apply the results that will be presented in the next sections.

\section{System Throughput Analysis and Optimization} \label{sec:thr}
In this section, we focus on the throughput analysis of the system depicted in Fig.~\ref{fig:model}. More explicitly, we intend to derive and optimize the weighted sum of the throughput $T_S$ at the helper $S$ and the throughput $T_U$ seen by the user $U$.
Define the weighted sum throughput by $w T_S + (1-w) T_U$ with $w \in [0,1]$, where $w=1$ and $w=0$ represent the cases where we are only interested in the helper $S$ or in the user $U$, respectively. 

The average service rate of the caching helper $S$ to its destination $D$ is 
\begin{equation}
\mu = q_S(1-q_U)p_{SD/S}+q_S q_U a p_{SD/S,DC}.
\end{equation}
Before proceeding to the next step, we provide the formal definition of queue stability as follows \cite{Szpankowski:stability}.

\begin{definition}
Denote by $Q_i^t$ the length of queue $i$ at the beginning of time slot $t$. The queue is said to be \emph{stable} if
$\lim_{t \rightarrow \infty} {Pr}[Q_i^t < {x}] = F(x)$ and $\lim_{ {x} \rightarrow \infty} F(x) = 1$.
\end{definition}

Although we do not make explicit use of this definition we use its corollary consequence which is Loynes' theorem~\cite{b:Loynes} that states that if the arrival and service processes of a queue are strictly jointly stationary and the average arrival rate is less than the average service rate, then the queue is stable. In the considered network model, the queue at $S$ is stable if and only if $\lambda < \mu$. The stability of a queue implies finite queuing delay. By adding this constraint we can achieve finite queueing delay. The stability at $S$ implies also that the packets that will enter the queue at $S$ eventually will be transmitted successfully to the destination $D$.

Denote $T_S$ the throughput on the link $S \rightarrow D$,
depending on whether the queue is stable, we have $T_S=\lambda$ if the queue is stable, or $T_S= \mu$ if the queue is unstable. Thus, 
\begin{equation} \label{eq:TS}
T_S= \mathbbm{1}(\lambda < \mu) \lambda + (1-\mathbbm{1}(\lambda < \mu) ) \mu, 
\end{equation}
where $\mathbbm{1}(.)$ is the indicator function.

The throughput seen by $U$ depends on the status of the queue at $S$:
\begin{itemize}
	\item When the queue at $S$ is empty with probability $\mathbb{P} (Q=0)$, $U$ requests for a file with probability $q_U$, which will be directed to $S$ with probability $q_C$. Then the throughput of $U$ is $q_U q_C p_h p_{SU/S}+ q_U (1-q_C) \alpha p_{DC/DC}$.
	\item When the queue at $S$ is non-empty with probability $\mathbb{P} (Q \neq 0)$,
	$U$ is active with probability $q_U$ and $S$ is available with probability $1-q_S$, then $U$ will seek for content from $S$ with probability $q_C$ or from $DC$ with probability $1-q_C$. In the first case, the achieved throughput by $U$ is $(1-q_S) q_U q_C p_h p_{SU/S}$.
	In the latter case, the throughput seen by $U$ is $(1-q_S ) q_U (1-q_C) \alpha p_{DC/DC}$.
	If $S$ is not available because it is transmitting a packet to $D$, which is with probability $q_S$, the throughput seen by $U$ is $q_S q_U \alpha p_{DC/S,DC}$.
\end{itemize}
Combining the above cases, the achieved throughput by $U$ is
\begin{align} \label{eq:TU_general}
T_U=&q_S \mathbb{P} (Q \neq 0) q_U \alpha p_{DC/S,DC}+ \left[1-q_S \mathbb{P} (Q \neq 0) \right] \!q_U \!\nonumber \\ 
&\cdot\left[ q_C p_h p_{SU/S} + (1-q_C p_h) \alpha p_{DC/DC} \right].
\end{align}
Depending on whether the queue at $S$ is stable, the expression of $T_U$ is different. 
In the remainder of this section, we will analyze and optimize the weighted sum throughput $w T_S + (1-w) T_U$ in the cases with stable and unstable queue at $S$, respectively.

\subsection{The Queue at $S$ is Stable}
When the queue at the caching helper $S$ is stable, 
the probability that the queue size $Q$ is not empty is given by
\begin{equation} \label{eq:prQempty}
\mathbb{P} (Q \neq 0) = \frac{\lambda}{\mu}=\frac{\lambda}{q_S(1-q_U)p_{SD/S}+q_S q_U \alpha p_{SD/S,DC}}.
\end{equation}
After replacing \eqref{eq:prQempty} in \eqref{eq:TU_general} we have the following expression for the throughput at $U$ 
\begin{align} \label{eq:TU}
T_U =&\frac{q_U \alpha \lambda p_{DC/S,DC}}{(1-q_U)p_{SD/S}+q_U \alpha p_{SD/S,DC}} + \nonumber \\ &\frac{(1-q_U)p_{SD/S}+q_U \alpha p_{SD/S,DC}-\lambda}{(1-q_U)p_{SD/S}+q_U \alpha p_{SD/S,DC}}q_U A,
\end{align}
where $A= \alpha p_{DC/DC} + q_C p_h (p_{SU/S} - \alpha p_{DC/DC})$. 
Note that \eqref{eq:TU} is independent of $q_S$ when the queue at $S$ is stable.

We define the following optimization problem, which aims at optimizing the probabilities $q_S$ and $q_C$ such that the weighed sum throughput is maximized.
\begin{eqnarray}
\underset{q_S, q_C} {\max} ~&&  w \lambda + (1-w) T_U (q_S, q_C)\\
\mbox{s.t.} 
&& \frac{\lambda}{(1-q_U)p_{SD/S}+q_Uap_{SD/S,DC}} \leq q_S \leq 1 \nonumber \\
&&   0 \leq q_C \leq 1 . \nonumber
\end{eqnarray}
The first constraint ensures stability at the helper's queue.
If $w=1$, the result of the previous optimization problem is $\lambda$,
thus, the objective function is independent of $q_C$ and $q_S \in [\frac{\lambda}{(1-q_U)p_{SD/S}+q_U \alpha p_{SD/S,DC}},1]$. If $w \neq 1$, the objective function is linear with respect to $q_C$; as stated in the previous section it does not depend directly on $q_S$. If $p_{SU/S} > \alpha p_{DC/DC} $ the objective function is an increasing function of $q_C$, thus, the maximum is achieved by $q_C^{*}=1$. 
If $p_{SU/S} < \alpha p_{DC/DC}$, then the objective function is a decreasing function of $q_C$, thus $q_C^{*}=0$.
The coefficient $p_{SU/S}-\alpha p_{DC/DC}$ is an indication of the channel between the helper and the user, the availability of $DC$, and the channel between the $DC$ and $U$. \textit{Note that the choice of $q_S$ does not affect the optimal solution when it lies in the interval that keeps the queue stable.}

\subsection{The Queue at $S$ is Unstable}

If the queue at $S$ is unstable means that the arrival rate is greater than the service rate. So, it is equivalent to disregard the bursty traffic and consider a saturated queue. In a network when the queue is unstable, a packet dropping policy can be applied in order to stabilize the system. However, if the system can be stabilized by packet dropping, then the results for the stable queue can still hold in this case with an arrival rate $\lambda^{'} < \lambda$ where, $\lambda^{'}$ is the arrival rate after applying the packet dropping.
If the queue at $S$ is unstable, the throughput seen by the user $T_U^{'}$ is given by
\begin{align} \label{eq:TU'}
T_U^{'}= & q_S q_U \alpha p_{DC/S,DC}+ \nonumber \\
&(1-q_S) q_U \left[ q_C p_h p_{SU/S} + (1-q_C p_h) \alpha p_{DC/DC} \right].
\end{align}
Then the weighted sum throughput optimization problem becomes
\begin{eqnarray} \label{eq:optunstable}
\underset{q_S, q_C} {\max} ~&& f(q_S, q_C) = w \mu + (1-w) T_U^{'} (q_S, q_C)\\
\mbox{s.t.} 
&& q_S, q_C \in [0,1] \nonumber.
\end{eqnarray}
The objective function can be re-written as
\begin{equation}
f(q_S, q_C)= q_C (1-q_S) B_1+q_S B_2 + (1-w) q_U \alpha p_{DC/DC},
\end{equation}
where $B_1=(1-w)q_U p_h (p_{SU/S} - \alpha p_{DC/DC})$ and $B_2=\alpha q_U (1-w) (p_{DC/S,DC}-p_{DC/DC})+p_{SD/S}(1-q_U)w+\alpha q_U p_{SD/S,DC}$.

The optimization problem stated in \eqref{eq:optunstable} is non-linear.
However, after applying the Karush Kuhn Tucker (KKT) conditions we can obtain the solution, which depends on the signs and the ordering between $B_1$ and $B_2$. Thus, for sake of presentation we omit the enumeration of the possible solutions based on $B_1$ and $B_2$. The possible optimal values of $q_C^*$ and $q_S^*$ that maximize the objective function are $q_C^* \in \{0,1/2,1\}$ and $q_S^* \in \{0,1/2,1\}$. In Section \ref{sec:results}, we will provide the numerical solutions in some specific cases.

\section{Delay Analysis} \label{sec:delay}
Apart from the system throughput, the delay experienced by the user is another critical metric for the performance of wireless caching systems with delay-sensitive applications.
In this section, we study the delay experienced by the user $U$
from the time that it requests for external content until the time that the content is received. We do not take into account the delay to access the requested content that is stored in its cache since, this can occur instantaneously.

Recall that we consider slotted transmission in this work. The delay for the user to received a requested file is one slot if one of the following events happen:
\begin{itemize}
	\item The user can locate the requested file within the helper's storage, the helper is available to assist the user, and the transmission from the helper to the user is successful.
	\item The requested content is not available at the helper, then the user within the same slot will be re-directed to the data center. The data center is available and the transmission from the data center to the user is successful.
\end{itemize}
In other cases, for instance, when the helper's cache contains the requested file, but the transmission fails, then in the next time slot the user will request for the same file again from the helper if the helper is available to assist. Otherwise, the user will be re-directed to the data center.
If the content is not available at the helper, and the transmission from the data center fails, then in the next time slot the user will seek for its content directly from the data center.

Based on the the aforementioned cases and applying the regenerative method \cite{Walrand}, we will derive the delay experienced by the user when the queue at the helper is stable and unstable, respectively.

\subsection{The queue at $S$ is stable}
The average delay $D$ that the user experiences to receive a requested file is
\begin{equation} \label{eq:delay}
\begin{split}
D&=\left[1-q_S\mathbb{P} (Q \neq 0)\right][p_h q_C + p_h q_C(1-p_{SD/S})D_S+\\ 
&(1-p_h)\alpha p_{DC/DC}+(1\!-\!p_h)(1-\alpha p_{DC/DC})(1+D_{DC})+\\
&p_h(1\!-\!q_C)\alpha p_{DC/DC}+p_h(1\!-\!q_C)(1\!-\!\alpha p_{DC/DC})(1+\!D_{S})] \\
&+q_S\mathbb{P} (Q \neq 0) [\alpha p_{DC/S,DC}+(1-\alpha p_{DC/S,DC})(1+D)],
\end{split}
\end{equation}
where $\mathbb{P} (Q \neq 0)$ is given by \eqref{eq:prQempty}. $D_S$ and $D_{DC}$ are given by
\begin{equation} \label{eq:delayS}
\begin{split}
D_S&=q_C \left[1-q_S\mathbb{P} (Q \neq 0)\right] \left[p_{SD/S}+(1-p_{SD/S})(1+D_S) \right] \\
&+(1-q_C)\left[1-q_S\mathbb{P} (Q \neq 0)\right]\\
&\cdot[\alpha p_{DC/DC}+(1-\alpha p_{DC/DC})(1+D_S)] \\
&+q_S\mathbb{P} (Q \neq 0) \!\left[\alpha p_{DC/S,DC} + \!(1\!-\!\alpha p_{DC/S,DC})(1\!+\!D_S)\right]
\end{split}
\end{equation}
and
\begin{equation} \label{eq:delayDC}
\begin{split}
D_{DC}=&\big[\alpha q_S\mathbb{P} (Q \neq 0) p_{DC/S,DC} \\
&+\alpha(1-q_S\mathbb{P} (Q \neq 0) )p_{DC/DC} \big]^{-1}.
\end{split}
\end{equation}
After some manipulations, \eqref{eq:delayS} becomes
%{\scriptsize
\begin{equation} \label{eq:delayS-simple}
\begin{split}
D_S=&\{q_S\mathbb{P} (Q \neq 0) \alpha p_{DC/S,DC} \\+& (1-q_S\mathbb{P} (Q \neq 0))[q_C p_{SD/S}+(1-q_C)\alpha p_{DC/DC}]\}^{-1}.
\end{split}
\end{equation}
%}

The average delay $D$ in \eqref{eq:delay} after some manipulations can be written as
\begin{equation} \label{eq:delay-simple}
D=\frac{1+\left[1-q_S\mathbb{P} (Q \neq 0)\right]F(D_S,D_{DC})}{1-q_S\mathbb{P} (Q \neq 0)(1-\alpha p_{DC/S,DC})},
\end{equation}
where
\begin{equation}\label{eq:F}
\begin{split}
&F(D_S,D_{DC})\\
= &p_h q_C(1-p_{SD/S})D_S + (1-p_h)(1-\alpha p_{DC/DC})D_{DC}\\
&+p_h(1-q_C)(1-\alpha p_{DC/DC})D_{S}.
\end{split}
\end{equation}

\subsection{The queue at $S$ is unstable}
If the queue at the helper $S$ is unstable (or the helper has saturated traffic), then the average delay at the user can be found by replacing $\mathbb{P} (Q \neq 0)=1$ to the previous expressions. Thus, we have
\begin{equation} \label{eq:delay-unstable}
D=\frac{1+(1-q_S)F(D_S,D_{DC})}{1-q_S(1-\alpha p_{DC/S,DC})},
\end{equation}
where $F(D_S,D_{DC})$ is given by \eqref{eq:F}. The expressions for $D_S$, $D_{DC}$
are given by
\begin{equation} \label{eq:delayS-simple-unstable}
\begin{split}
D_S=&\big[q_S \alpha p_{DC/S,DC} + \\
&(1-q_S)[q_C p_{SD/S}+(1-q_C)\alpha p_{DC/DC}]\big]^{-1},
\end{split}
\end{equation}
and
\begin{equation} \label{eq:delayDC-unstable}
\begin{split}
D_{DC}=\left[a\left[q_S p_{DC/S,DC} +(1-q_S)p_{DC/DC} \right]\right]^{-1}.
\end{split}
\end{equation}

\section{Numerical Results} \label{sec:results}
In this section, we provide numerical evaluation of the results presented in the previous sections. We assume $\eta_j=10^{-11}$ W, $\theta_1=\theta_2=0$ dB, and the path-loss exponent is $\gamma=4$. Transmission powers are $P_{tx}(S)=1$ mW and $P_{tx}(DC)=10$ mW.
The distances of links $S \rightarrow D$, $S \rightarrow U$, $DC \rightarrow D$ and $DC \rightarrow U$ are $r_{SD}=50$m, $r_{SU}=40$m, $r_{DCD}=100$m, and $r_{DCU}=80$m  respectively. From (\ref{eq:succprobSNR}) and (\ref{eq:succprobSINR}) we obtain the following success probabilities $p_{SD/S}=0.939$, $p_{SD/S,DC}=0.578$, $p_{SU/S}=0.975$, $p_{DC/DC}=0.96$ and $p_{DC/S,DC}=0.369$. Note that the values of the success probabilities do not have specific implications in the numerical evaluation. Our general remarks in this section are valid for other parameter values as well.

We also assume that the total number of available files is $10000$, the shape parameter of the Zipf law is $\delta=0.5$, and the cache size at the source is $M_S=2000$. Furthermore, the data center is available with probability $\alpha=0.7$.
The helper and the user apply the MPC policy, as described in Section \ref{sec:model}.

\subsection{Maximum Weighted Sum Throughput: The effect of the arrival rate $\lambda$}
\begin{figure}[ht]
\centering
\includegraphics[scale=0.5]{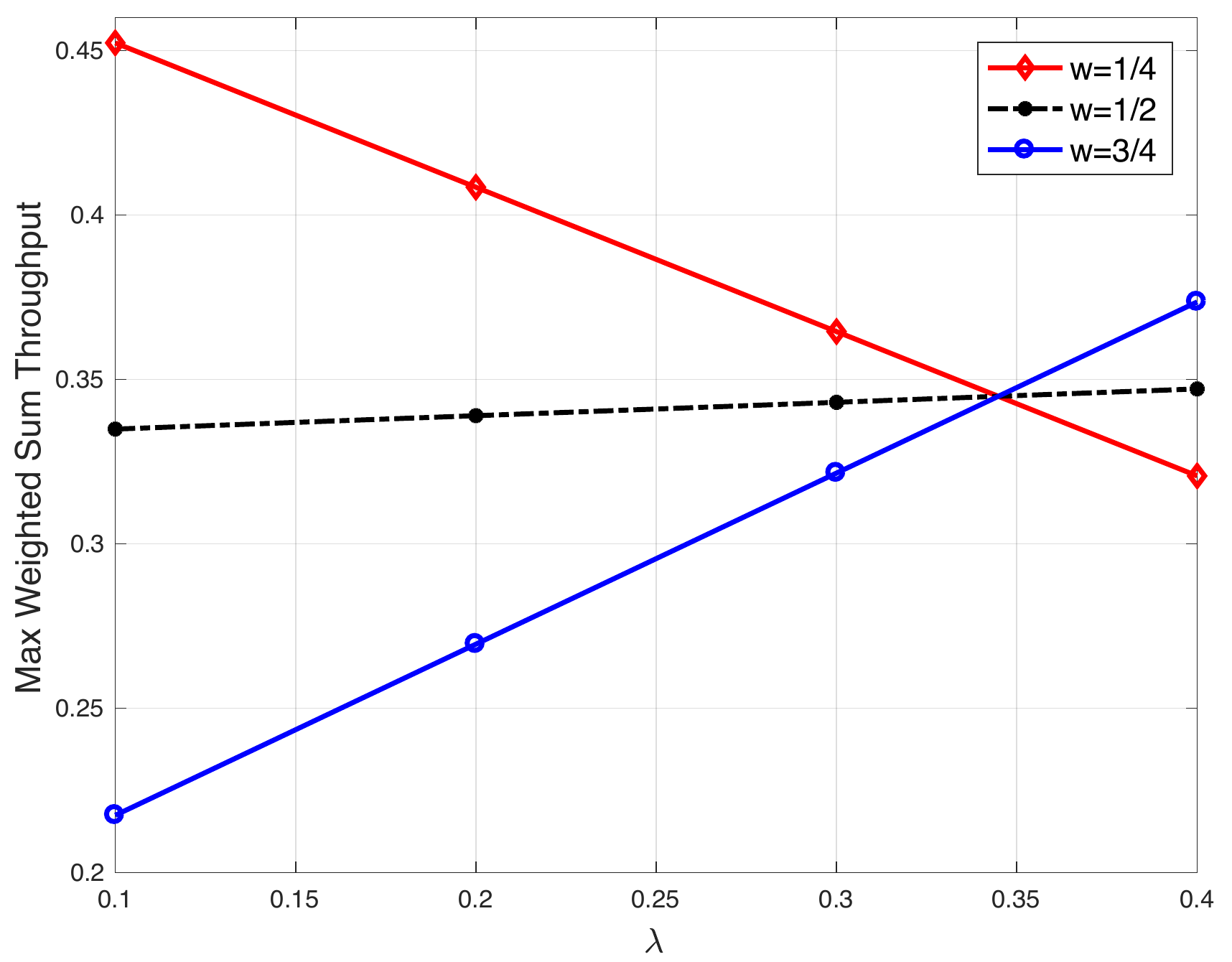}
\caption{The maximum weighted sum throughput vs. $\lambda$.}
\centering
\label{fig:thrvslambda}
\end{figure}

We consider the scenario where the capacity of the cache user is $M_U=200$. Thus, after replacing the values into \eqref{eq:phMPC} and \eqref{eq:qU}, we obtain $p_h = 0.31$ and $q_U = 0.86$. In Fig. \ref{fig:thrvslambda}, the weighted sum throughput vs. the arrival rate at the helper is presented for three different values of $w$. $w=1/4$ represents the case when $T_U$ is more important than $T_S$. Thus, the maximum weighted sum throughput is a decreasing function of $\lambda$ for $w=1/4$. For $w=1/2$ and $w=3/4$, the maximum weighted sum throughput is an increasing function of $\lambda$.

The maximum weighted sum throughput is always achieved by $q_{C}^*=1$ for any $w$ and $\lambda$ under this specific setup. However, the values of $q_S^*$ that achieve the maximum weighted sum throughput are different, as presented in Table \ref{tab:qsmax_thrvslambda}. Note that these values are independent of $w$. As expected, as $\lambda$ increases, $q_S^*$ increases in order to keep the queue at $S$ stable.

\begin{table}[ht]
\begin{center}
	\renewcommand{\arraystretch}{1.2}
	\caption{The values of $q_S^*$ that achieve the maximum weighted sum throughput presented in Fig. \ref{fig:thrvslambda}.}
\begin{tabular}{|c|c|c|c|c|}
\hline
$\lambda$ & 0.1 & 0.2 & 0.3 & 0.4 \\\hline
$q_S^*$ & 0.209639 & 0.41927 & 0.628917 & 0.838556 \\\hline
\end{tabular}
\label{tab:qsmax_thrvslambda}
\end{center}
\end{table}

For the case where the queue is unstable, when the arrival rate is greater than the average service rate, the maximum weighted sum throughput and the values of $q_S^*$ and $q_C^*$ are presented in Table \ref{tab:unstable-max1}.

\begin{table}[ht]
\begin{center}
		\renewcommand{\arraystretch}{1.2}
		\caption{The maximum weighted sum throughput and the values of $q_S^*$ and $q_C^*$ for the case with unstable queue.}
\begin{tabular}{|c|c|c|c|}
\hline
$w$ & Max & $q_S^*$ & $q_C^*$\\\hline
1/4 & 0.496229 & 0 & 1 \\ \hline
1/2 & 0.350238 & 1& 0  \\ \hline
3/4 & 0.413624 & 1 & 0\\ \hline
\end{tabular}
\label{tab:unstable-max1}
\end{center}
\end{table}

\subsection{Maximum Weighted Sum Throughput: The effect of the capacity of the cache at the user $M_U$}

Here, we study the effect of the storage capacity at $M_U$ on the maximum weighted sum throughput. We consider several values for $M_U$ and their connection with $p_h$ and $q_U$ as summarized in Table \ref{tab:MUqUph}.

\begin{table}[ht]
\begin{center}
		\renewcommand{\arraystretch}{1.2}
		\caption{The values of $q_U$ and $p_h$ obtained by different values of the cache capacity $M_U$ at the user.}
\begin{tabular}{|c|c|c|c|c|c|c|}
\hline
$M_U$ & 100 & 200 & 400 & 800 & 1600 & 2000 \\\hline
$q_U$ & 0.91 & 0.86 & 0.81 & 0.72 & 0.6 & 0.56  \\ \hline
$p_h$ & 0.35 & 0.31 & 0.25 & 0.17 & 0.05 & 0 \\ \hline
\end{tabular}
\label{tab:MUqUph}
\end{center}
\end{table}

We observe that as $M_U$ increases, $q_U$ decreases because it is more likely for the user to find the requested content in its cache. Since we have assumed the MPC scheme, as $M_U$ increases then $p_h$ decreases.

\begin{figure}[ht]
\centering
\includegraphics[scale=0.53]{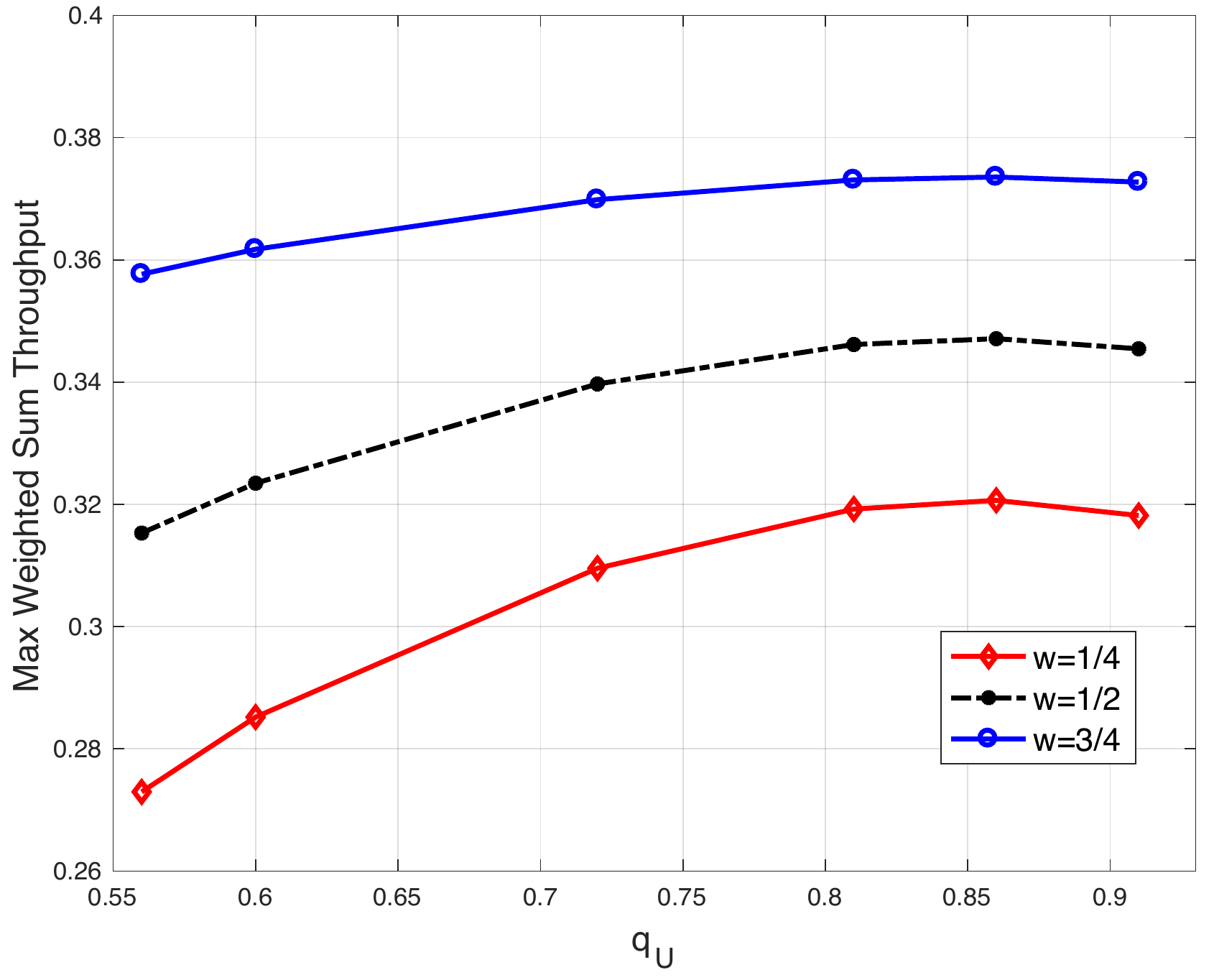}
\caption{The maximum weighted sum throughput vs. $q_U$ for $\alpha = 0.7$ and $\lambda=0.4$.}
\centering
\label{fig:thrvsqu}
\end{figure}

In Fig. \ref{fig:thrvsqu} the maximum weighted sum throughput vs. $q_U$ is presented with $\alpha = 0.7$ and $\lambda=0.4$, when the queue at $S$ is stable. We used the values of the $q_U$ described in Table \ref{tab:MUqUph}.
As $q_U$ increases, due to the decrease of the storage capacity, the improvement of the maximum weighted sum throughput is more profound for the case with $w=1/4$, where the throughput seen at $U$ is more important. This increase is also explained by the fact that the traffic inside the network increases as $q_U$ increases.

The values of $q_S^*$ and $q_C^*$ that achieve the maximum weighted sum throughput are given in Table \ref{tab:max2}. As expected, when the user has higher probability to request for external content, the transmission probability by the source to its destination increases in order to sustain stability. This is because the source is more likely to assist the user, $q_C^*=1$ for $q_U>0.56$, and also to overcome the interference caused by the transmission from the data center to the user. Observe that the only case that we have $q_C^*=0$ is when $M_U=2000$ thus, $p_h=0$, in this case the helper cannot actually assist the user with content.

\begin{table}[ht]
\begin{center}
		\renewcommand{\arraystretch}{1.2}
		\caption{The values of $q_S^*$ and $q_C^*$ that achieve the maximum weighted sum throughput presented in Fig. \ref{fig:thrvsqu} for different values of $q_U$.}
		
\begin{tabular}{|c|c|c|c|c|c|c|}
\hline
$q_U$ & 0.56 & 0.6 & 0.72 & 0.81 & 0.86 & 0.91 \\\hline
$q_S^*$ & 0.623 & 0.649 & 0.723 & 0.787 & 0.839 & 0.88 \\\hline
$q_C^*$ & 0 & 1 & 1 & 1 & 1 & 1 \\\hline
\end{tabular}
\label{tab:max2}
\end{center}
\end{table}

\begin{figure}[ht]
\centering
\includegraphics[scale=0.5]{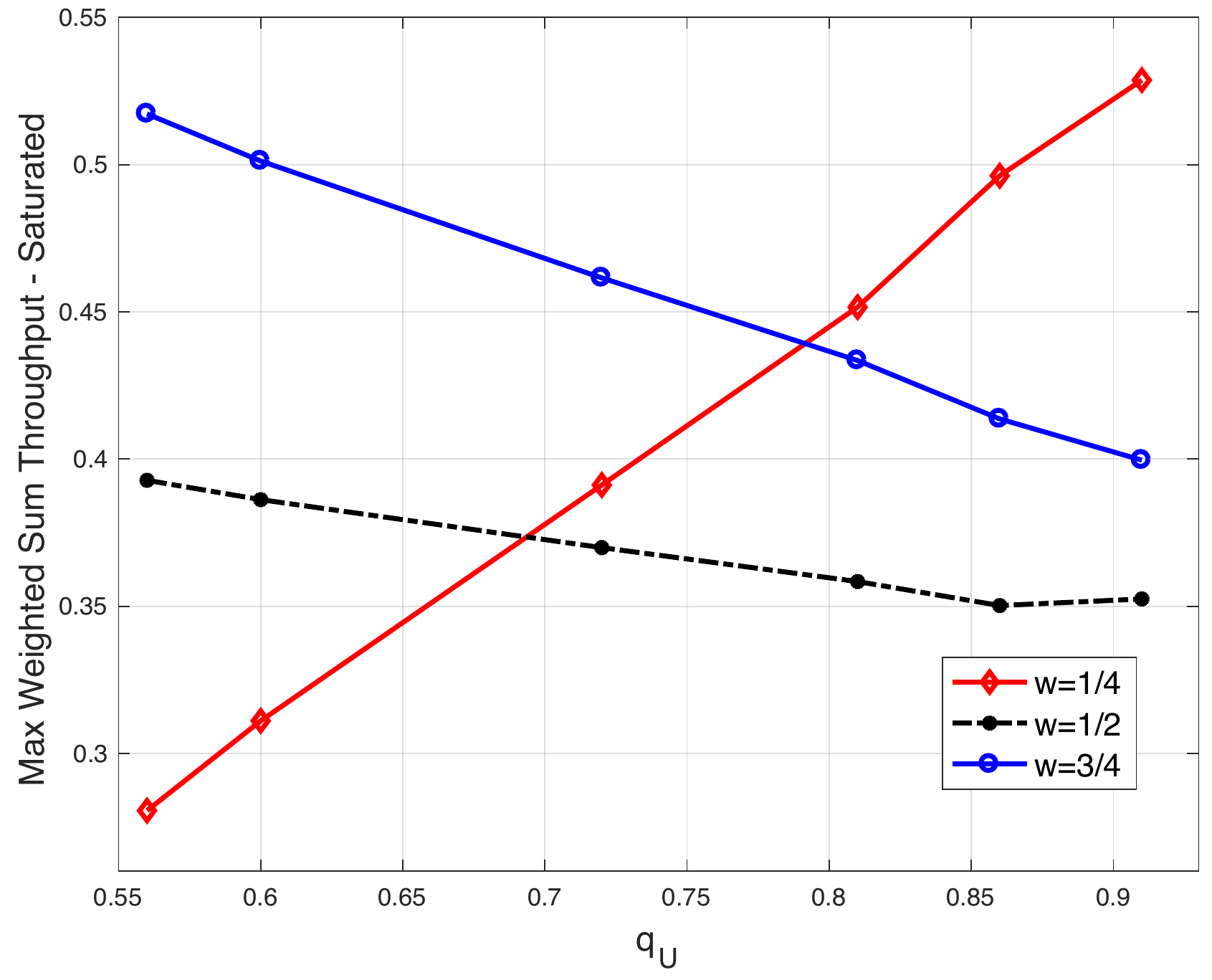}
\caption{The maximum weighted sum throughput vs. $q_U$ for $\alpha = 0.7$ when the queue at the helper is saturated/unstable.}
\centering
\label{fig:thrvsqu-sat}
\end{figure}

In Fig. \ref{fig:thrvsqu-sat} the maximum weighted sum throughput is presented when the queue at the source is unstable (or the source has saturated traffic) for $w=1/4, 1/2, 3/4$. The values of $q_S^*$ and $q_C^*$ that achieve the maximum are given in Tables \ref{tab:max32}, \ref{tab:max32}, and \ref{tab:max33} for $w=1/4$, $w=1/2$, and $w=3/4$ respectively.

For $w=1/2$ and $w=3/4$, the maximum weighted sum throughput is a decreasing function of $q_U$, since the saturated throughput achieved by $D$ is decreasing, due to the increase of requests by $U$.

With $w=1/4$, the throughput achieved by $U$ is more important, thus, the maximum weighted sum throughput increases as $q_U$ increases. However for $q_U=0.91$, we observe that the maximum is smaller than the one achieved with $q_U=0.839$; this is because of the increase in the traffic in the network the increased interference makes the performance slightly worse. The values of $q_S^*$ and $q_C^*$ that achieve the maximum are given in Table \ref{tab:max3}. For $q_U>0.56$ we observe that $q_S^*=0$ and $q_C^*=1$, this can be interpreted as it is more beneficial for the network performance that the helper will serve solely the user if available. For $q_U=0.56$, we have that $p_h=0$ thus, it is better to have a silent helper in order to allow an interference free transmission from the data center to the user when is requested.

\begin{table}[ht]
\begin{center}
		\renewcommand{\arraystretch}{1.2}
		\caption{The values of $q_S^*$ and $q_C^*$ that achieve the maximum weighted sum throughput presented in Fig. \ref{fig:thrvsqu-sat} for different $q_U$ when $w=1/4$.}
\begin{tabular}{|c|c|c|c|c|c|c|}
\hline
$q_U$ & 0.56 & 0.6 & 0.72 & 0.81 & 0.86 & 0.91 \\\hline
$q_S^*$ & 0 & 0 & 0 & 0 & 0 & 0 \\\hline
$q_C^*$ & 0 & 1 & 1 & 1 & 1 & 1 \\\hline
\end{tabular}
\label{tab:max3}
\end{center}
\end{table}

With $w=1/2$ and $w=3/4$, the values of $q_S^*$ and $q_C^*$ that achieve the maximum are given in Tables \ref{tab:max32} and \ref{tab:max33} for $w=1/2$ and $w=3/4$ respectively. The maximum is achieved by $q_S^*=1$ and $q_C^*=0$ for all the values of $q_U$ for the case $w=3/4$. In this case, the transmission from the source to its destination is more significant thus, it the user will not assist the possible requests by the user. This means that it will be more beneficial if the source $S$ will transmit with probability $1$ to its destination $D$ and the user $U$ will seek for content directly from $DC$. The same result we have for $w=1/2$ when $q_U \leq 0.86$. However, if $q_U=0.91$ then the maximum is achieved by $q_S^*=0$ and $q_C^*=1$.

\begin{table}[ht]
\begin{center}
		\renewcommand{\arraystretch}{1.2}
		\caption{The values of $q_S^*$ and $q_C^*$ that achieve the maximum weighted sum throughput presented in Fig. \ref{fig:thrvsqu-sat} for different $q_U$ when $w=1/2$.}
\begin{tabular}{|c|c|c|c|c|c|c|}
\hline
$q_U$ & 0.56 & 0.6 & 0.72 & 0.81 & 0.86 & 0.91 \\\hline
$q_S^*$ & 1 & 1 & 1 & 1 & 1 & 0 \\\hline
$q_C^*$ & 0 & 0 & 0 & 0 & 0 & 1 \\\hline
\end{tabular}
\label{tab:max32}
\end{center}
\end{table}

\begin{table}[ht]
\begin{center}
		\renewcommand{\arraystretch}{1.2}
		\caption{The values of $q_S^*$ and $q_C^*$ that achieve the maximum weighted sum throughput presented in Fig. \ref{fig:thrvsqu-sat} for different $q_U$ when $w=3/4$.}
\begin{tabular}{|c|c|c|c|c|c|c|}
\hline
$q_U$ & 0.56 & 0.6 & 0.72 & 0.81 & 0.86 & 0.91 \\\hline
$q_S^*$ & 1 & 1 & 1 & 1 & 1 & 1 \\\hline
$q_C^*$ & 0 & 0 & 0 & 0 & 0 & 0 \\\hline
\end{tabular}
\label{tab:max33}
\end{center}
\end{table}

\subsection{Average Delay Experienced by $U$ for seeking external content}
\begin{figure}[ht]
\centering
\includegraphics[scale=0.52]{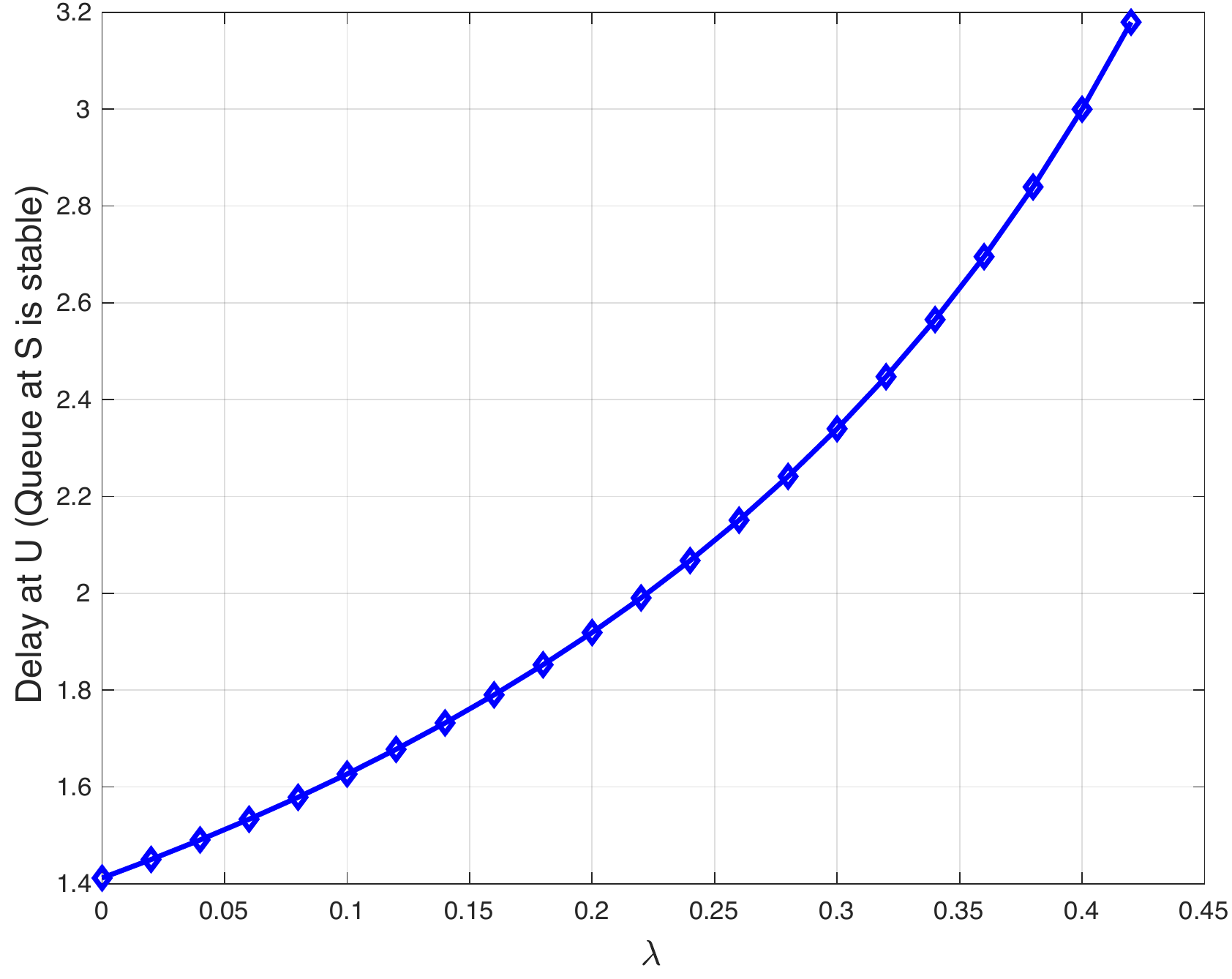}
\caption{The average delay vs. $\lambda$ for $\alpha = 0.7$ and $q_S=0.9$.} 
\centering
\label{fig:delayvslambda}
\end{figure}

\begin{figure}[ht]
\centering
 \subfigure[$\lambda=0.2$.]{
 \includegraphics[scale=0.5]{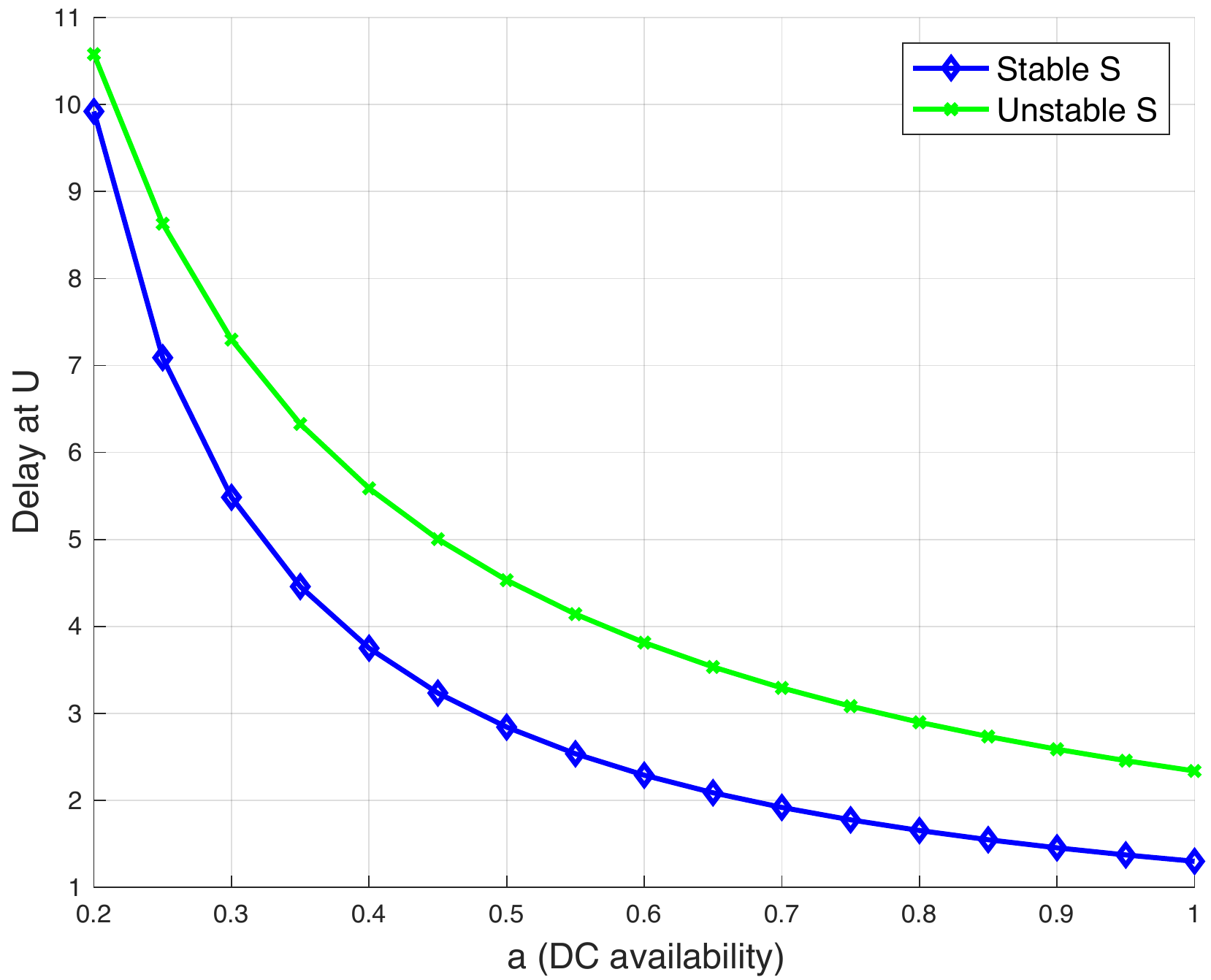}
 \label{fig:delayvsa02}
 }

  \subfigure[$\lambda=0.4$.]{
  \includegraphics[scale=0.5]{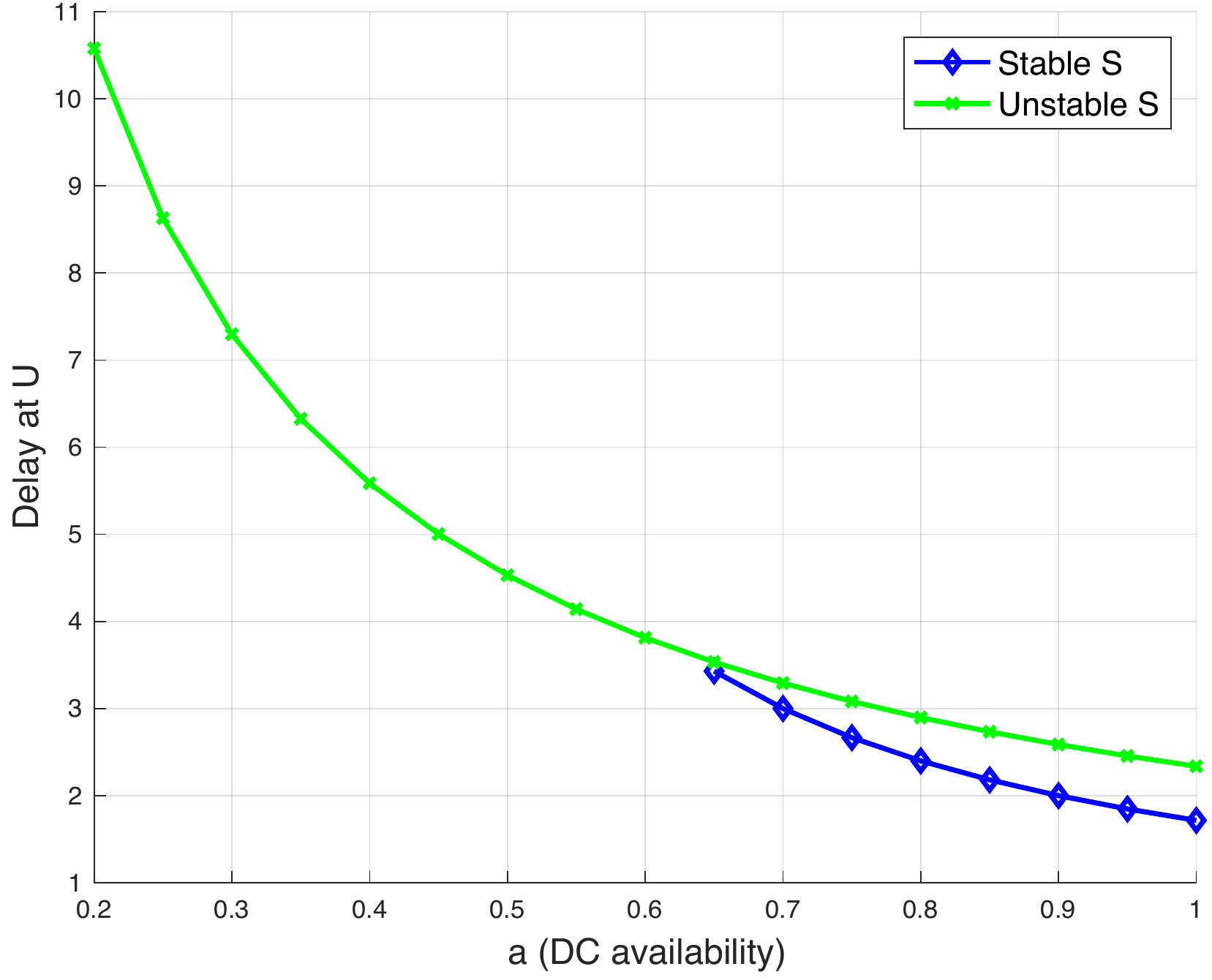}
  \label{fig:delayvsa04}
  }
  \caption{The average delay vs. $\alpha$ for $q_S=0.9$ for the cases of stable and unstable queue at $S$.}\label{fig:delayvsa}
\end{figure}

\begin{figure}[ht]
\centering
 \subfigure[$\lambda=0.2$.]{
 \includegraphics[scale=0.5]{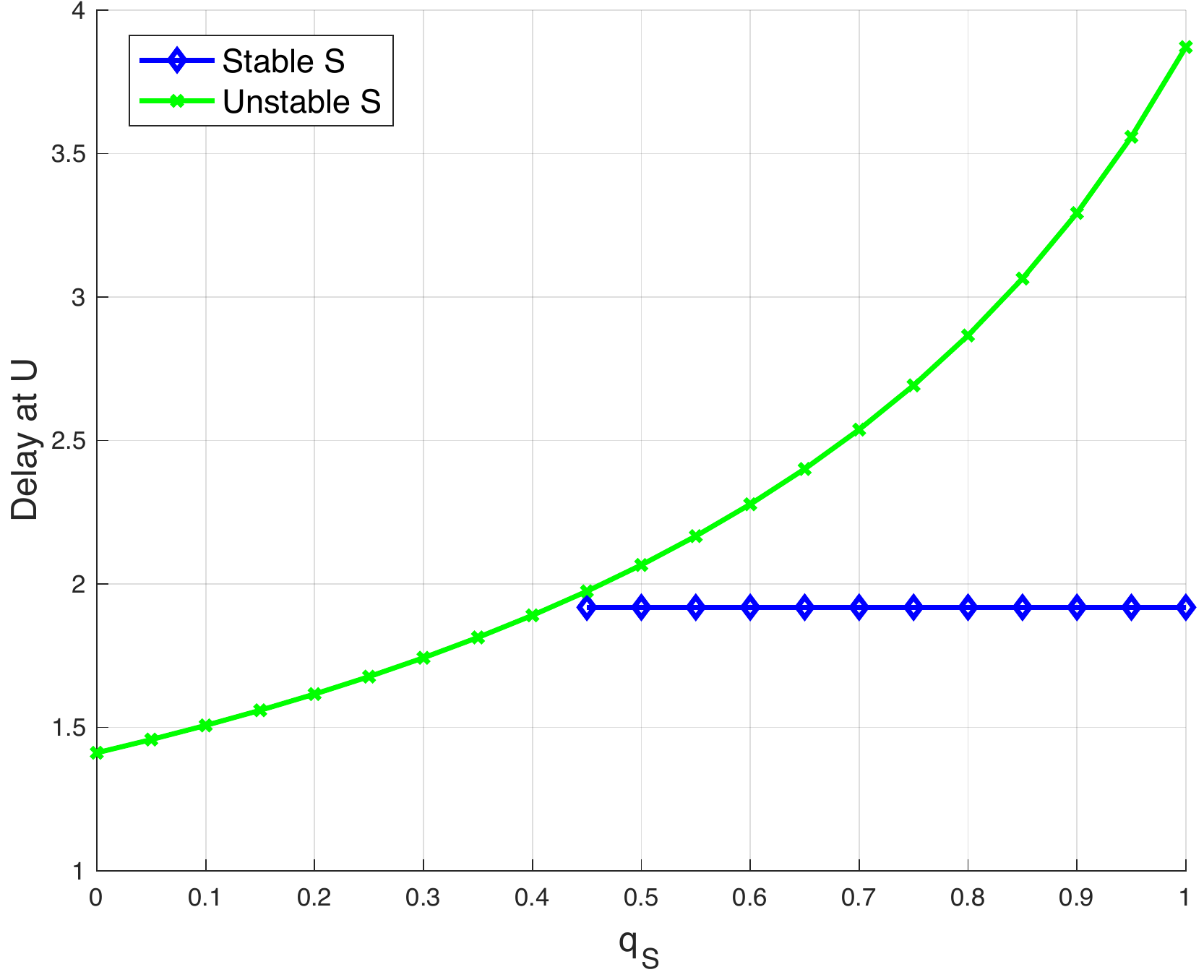}
 \label{fig:delayvsqs02}
 }

  \subfigure[$\lambda=0.4$.]{
  \includegraphics[scale=0.5]{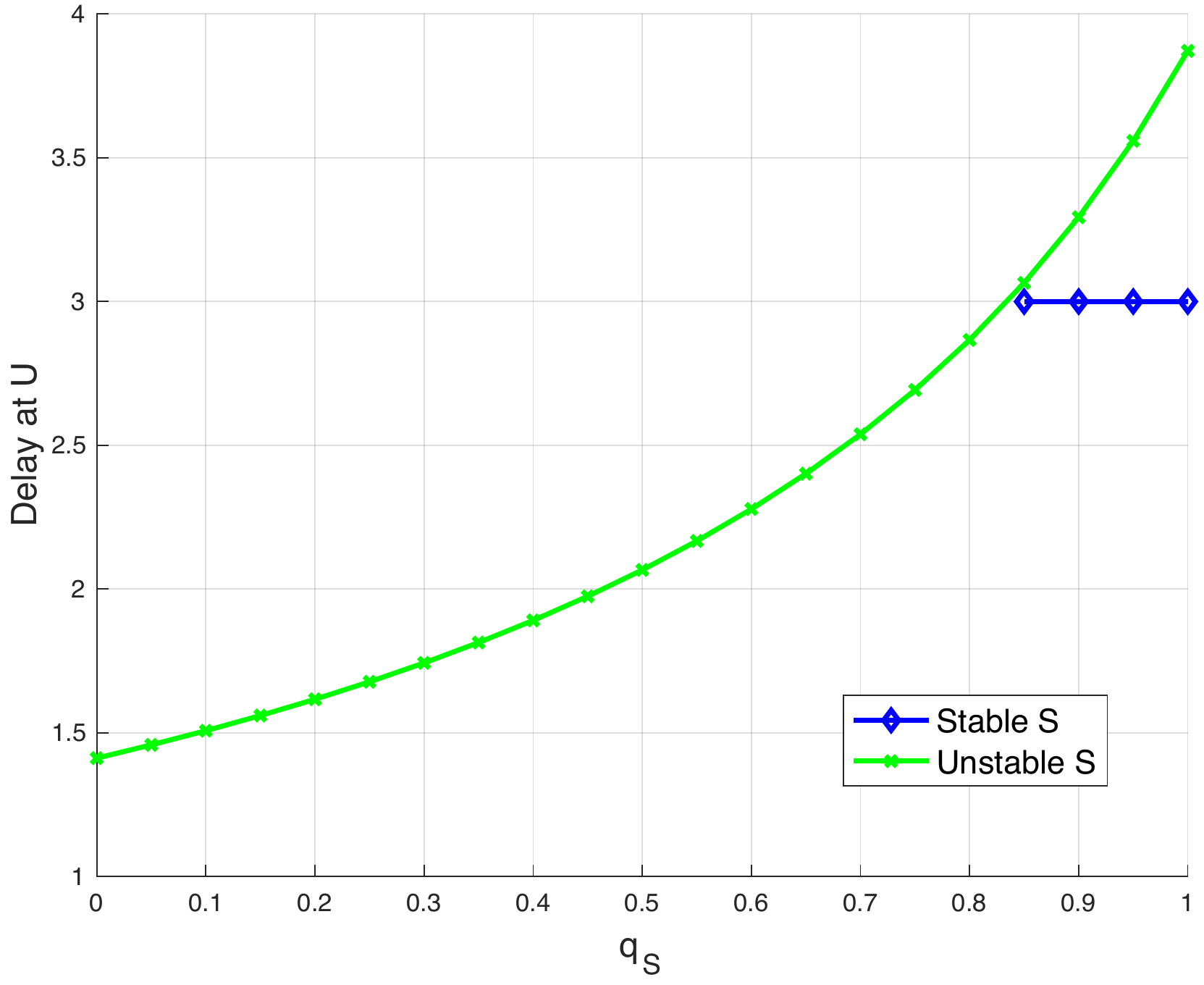}
  \label{fig:delayvsqs04}
  }
  \caption{The average delay vs. $q_S$ for $\alpha=0.7$ for the cases of stable and unstable queue at $S$.}\label{fig:delayvsqs}
\end{figure}

Here, we present the numerical results based on the analysis at Section \ref{sec:delay} regarding the delay experienced by the user to retrieve the requested external content. The Figs. \ref{fig:delayvslambda}, \ref{fig:delayvsa}, \ref{fig:delayvsqs} illustrate the effect of $\lambda$, $\alpha$ and $q_S$ on the average delay. In these plots we have $q_C=0.5$, $p_h=0.31$ and $q_U=0.86$.

In Fig. \ref{fig:delayvslambda}, the average delay versus the arrival rate at the helper is depicted for $\alpha = 0.7$ and $q_S=0.9$. We see that as the arrival rate at the helper increases, the delay increases non-linearly. The stopping value of $\lambda$ is obtained by the stability condition.

In Fig. \ref{fig:delayvsa}, we present the average delay versus the availability of the data center for both cases with stable and unstable queue at the helper. We consider two cases for the arrival rate, $\lambda=0.2$ and $\lambda=0.4$. The starting points at the figures for the stable case are obtained by the stability condition. 

As we observe, the delay is lower when the queue is stable since there are more chances for the user to find an available helper. However, in this case we have a requirement for a higher value of $\alpha$ in order to sustain a stable queue. This is profound in the case of $\lambda=0.4$.

In Fig. \ref{fig:delayvsqs}, we show the average delay versus $q_S$ for $\lambda=0.2$ and $\lambda=0.4$. We consider two cases regarding the queue at the helper. For the unstable queue we see an increasing trend on the delay when $q_S$ increases, which is expected since the chances of the user $U$ finding an available helper are smaller. The more interesting case is with the stable queue, where the delay at the user is insensitive to $q_S$. The reason behind this is that the product $q_S \mathbb{P} (Q \neq 0)$ is independent of $q_S$, meaning that the availability of the helper for constant $q_C$ does not depend on $q_S$ whenever the queue is stable.

For the case where $\lambda=0.4$, the minimum value of $q_S$ that can sustain a stable queue is higher than $\lambda=0.2$. This is because the helper needs to transmit more frequently to its own destination in order to stabilize its queue.

\begin{figure}[ht]
\centering
 \subfigure[$\lambda=0.2$.]{
 \includegraphics[scale=0.5]{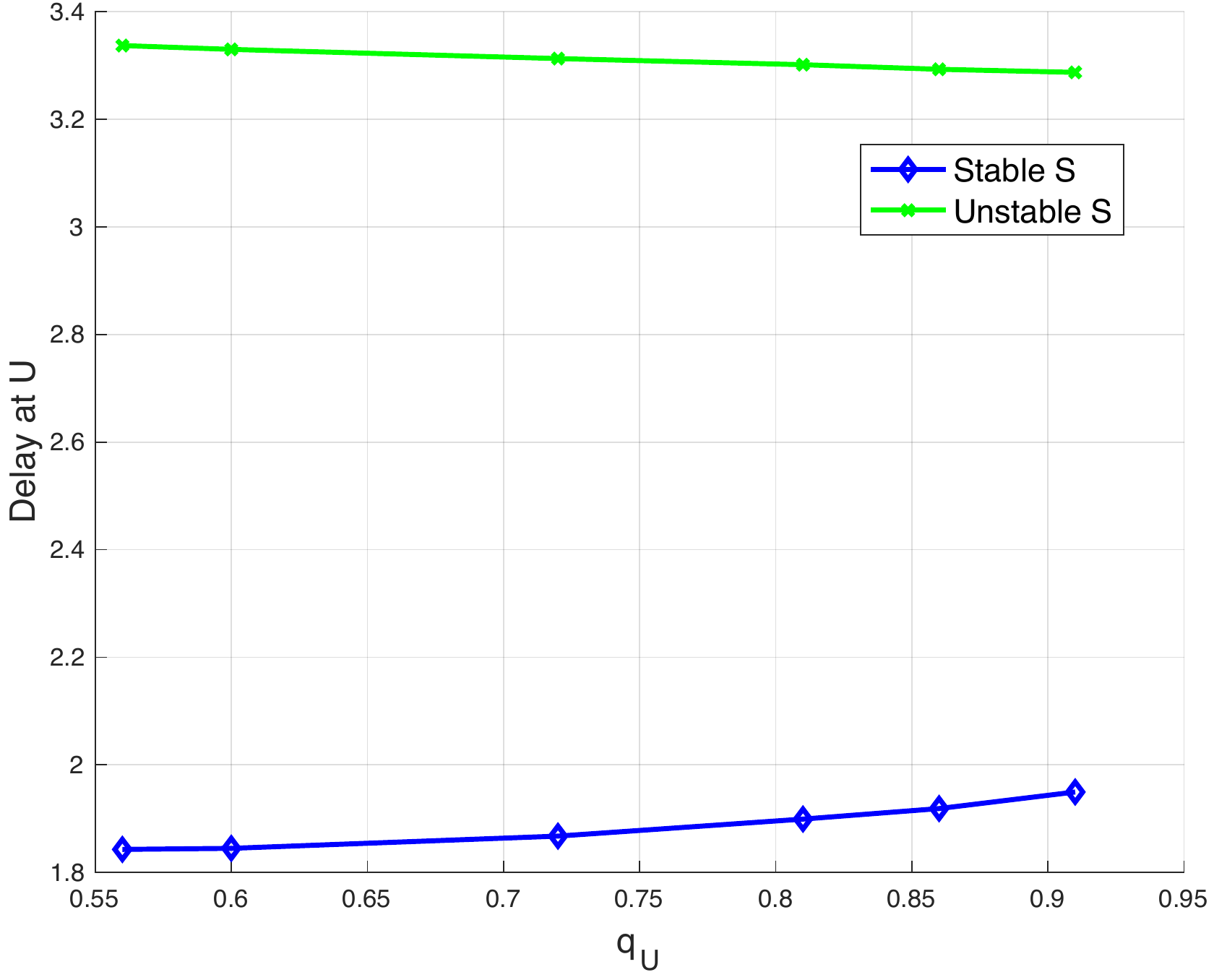}
 \label{fig:delayvsqu02}
 }
  \subfigure[$\lambda=0.4$.]{
  \includegraphics[scale=0.5]{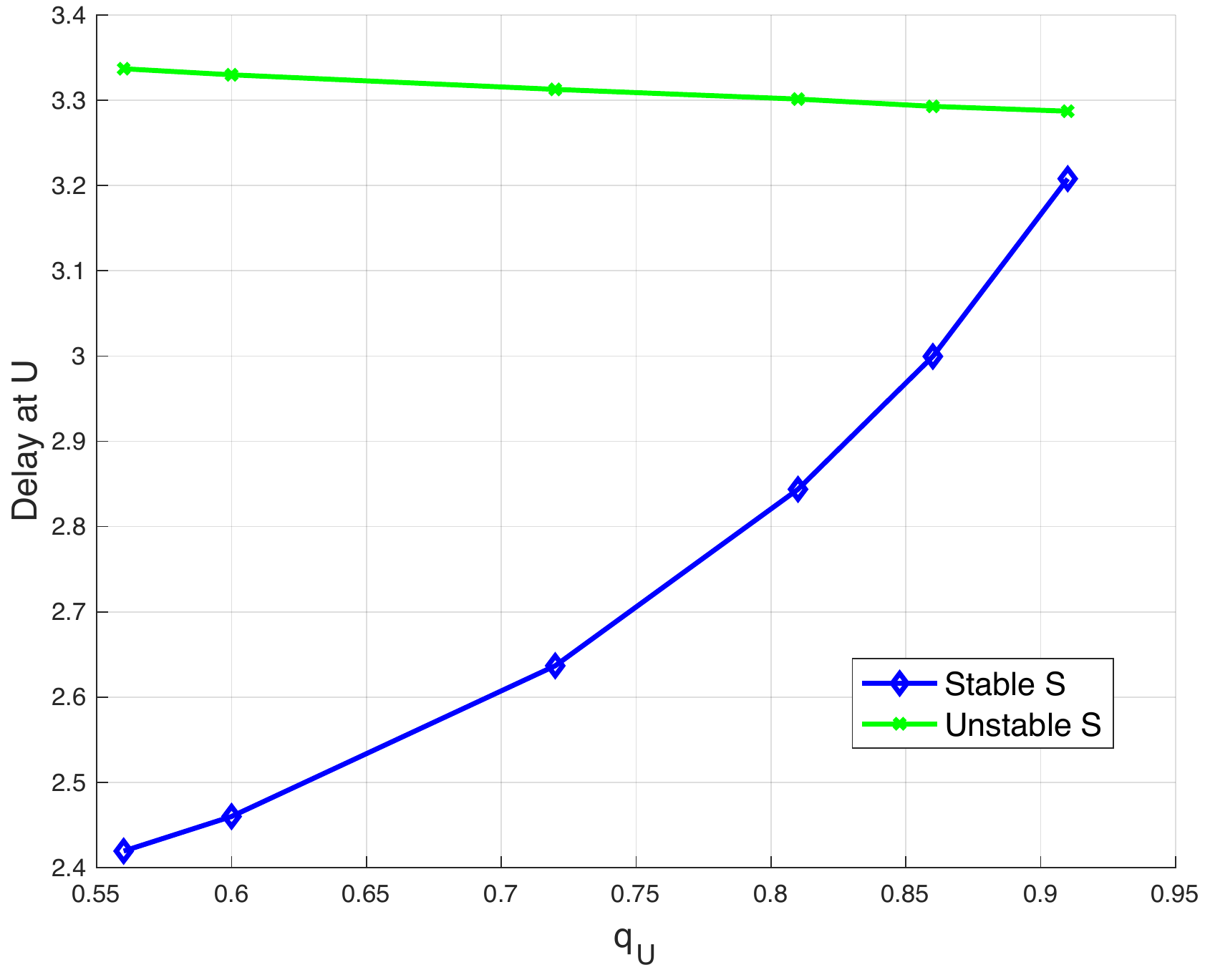}
  \label{fig:delayvsqu04}
  }
  \caption{The average delay vs. $q_U$ for $q_S=0.9$ for the cases of stable and unstable queue at $S$.}\label{fig:delayvsqu}
\end{figure}

Next, we will study the effect on delay of the storage capacity of user $U$.
The capacity $M_U$ of the user affects the request probability for external content, $q_U$. In Table \ref{tab:MUqUph} we provide the connection of the cache capacity with the request probability $q_U$ for several cases. In Fig. \ref{fig:delayvsqu}, we depict the average delay versus $q_U$ for two values of $\lambda$. In these figures we also illustrate the case where the queue at the helper is unstable.

When the congestion at the helper is low, $\lambda=0.2$, the effect of the storage of the user does not affect the delay significantly. When the congestion increases, for example in the case $\lambda=0.4$, then increasing the storage capacity at the user decreases the average delay. Recall that as $q_U$ increases the storage decreases, thus, it is more likely for the user to not be able to find the requested content in its own storage. 

When the queue is unstable, recall that the average delay does not depend on $q_U$ directly. However, the increase of $q_U$, because of the decrease of the storage, affects $p_h$ as we can also see by Table \ref{tab:MUqUph}. This explains the slight decrease at the delay due to the increase of $p_h$.

\section{Conclusions} \label{sec:conclusions}
In this work, the effect of bursty traffic and random availability of caching helper in a wireless caching system was investigated. We studied the throughput of a wireless caching system consisting of a caching helper with a dedicated user and another non-dedicated user in proximity whose requests shall be served either by the caching helper or by a data center. For the purpose of throughput maximization, we optimized the request probability of the user to be served by the helper and the probability that the helper will transmit information to its dedicated destination. In addition, we derived the average delay experienced by the user from the time that its request is placed until the time the content is received. 

Our results provide fundamental insights in the throughput and delay behavior of helper-based wireless caching systems, which are essential for further investigation of this topic in larger topologies. A future direction of this work lies in the case where multiple users are competing for assistance from a set of sources.

\bibliographystyle{IEEEtran}
\bibliography{bib_ref}
\end{document}